\documentclass[numberedappendix,apj]{emulateapj}
\usepackage{apjfonts}

\begin{document}
\title{The Fundamental Plane for \lowercase{$z=0.8-0.9$} cluster galaxies}

\author{Inger J{\o}rgensen\altaffilmark{1},
Kristin Chiboucas\altaffilmark{1},
Kathleen Flint\altaffilmark{1,2},
Marcel Bergmann\altaffilmark{3},
Jordi Barr\altaffilmark{4}, 
Roger Davies\altaffilmark{4}
}

\altaffiltext{1}{Gemini Observatory, 670 N.\ A`ohoku Pl., Hilo, HI 96720; 
ijorgensen@gemini.edu, kchibouc@gemini.edu}
\altaffiltext{2}{Currently at State University of New York at Stony Brook, The Reinvention Center,
Stony Brook, NY 11794; kathleen.flint@stonybrook.edu}
\altaffiltext{3}{Gemini Observatory, La Serena, Chile; mbergmann@gemini.edu}
\altaffiltext{4}{Department of Astrophysics, University of Oxford, Keble Road, 
Oxford OX1 3RH, UK; jmb@astro.ox.ac.uk, rld@astro.ox.ac.uk}

\submitted{Accepted for publication in  Astrophysical Journal Letters}

\begin{abstract}
We present the Fundamental Plane (FP) for 38 early-type galaxies 
in the two rich galaxy clusters RXJ0152.7--1357 ($z=0.83$) and 
RXJ1226.9+3332 ($z=0.89$), reaching a limiting magnitude of 
$M_B =-19.8\,{\rm mag}$ in the rest frame of the clusters.
While the zero point offset of the FP for 
these high redshift clusters relative to our low redshift 
sample is consistent with passive evolution with a formation 
redshift of $z_{\rm form}\approx 3.2$, the FP
for the high redshift clusters is not only shifted as 
expected for a mass-independent $z_{\rm form}$, but 
rotated relative to the low redshift sample. Expressed as a 
relation between the galaxy masses and the mass-to-light
ratios the FP is significantly steeper for the 
high redshift clusters than found at low redshift.
We interpret this as a mass dependency of the star formation 
history, as has been suggested by other recent studies. 
The low mass galaxies ($10^{10.3}\,M_{\sun}$) have 
experienced star formation as recently as $z\approx 1.35$ 
(1.5 Gyr prior to their look back time), while galaxies with 
masses larger than $10^{11.3}\,M_{\sun}$ had their last major 
star formation episode at $z > 4.5$.
\end{abstract}


\section{Introduction}

The Fundamental Plane (FP) for elliptical (E) and lenticular (S0) galaxies
is a key scaling relation, which relates the effective radii, the
mean surface brightnesses and the velocity dispersions in a 
relation linear in log-space (e.g., Dressler et al.\ 1987;
Djorgovski \& Davis 1987; J\o rgensen et al.\ 1996, hereafter JFK1996). 
The FP can be interpreted as a relation between the galaxy masses and
their mass-to-light (M/L) ratios. For low redshift cluster galaxies the FP 
has very low internal scatter, e.g.\ JFK1996.
It is therefore a powerful tool for studying the 
evolution of the M/L ratio as a function of redshift (e.g., J\o rgensen et al.\ 1999;
Kelson et al.\ 2000; van de Ven et al.\ 2003;
Gebhardt et al.\ 2003; Wuyts et al.\ 2004; Treu et al.\ 2005; 
Ziegler et al.\ 2005).
These authors all find that the FP at 
$z$=0.2--1.0 is consistent with passive evolution of the stellar
populations of the galaxies, generally with a formation 
redshift $z_{\rm form} > 2$.  Most previous studies
of the FP at $z$=0.2--1.0 cover fairly small samples
of galaxies in each cluster and are limited to a narrow range in
luminosities and therefore in masses, making it very difficult
to detect possible differences in the FP slope.
A few recent studies indicated a steepening of the FP 
slope for $z \sim 1$ galaxies
(di Serego Alighieri et al.\ 2005; van der Wel et al.\ 2005; Holden et al.\ 2005).
These studies and studies of the K-band luminosity function (Toft et al.\ 2004)
and the red sequence (de Lucia et al.\ 2004) at $z\approx 0.8-1.2$ suggest
a mass dependency of the formation epoch.

We present the FP for two galaxy clusters
RXJ0152.7--1357 at $z=0.83$ and RXJ1226.9+3332 at $z=0.89$. Our samples
reach apparent $i'$-band magnitudes of 22.5--22.8 mag, equivalent 
to an absolute magnitude of $M_B = -19.8\,{\rm mag}$ in the rest
frame of the clusters.  No other published samples suitable for 
studies of the cluster galaxy FP at $z>0.8$ go this deep.
Our study of these two clusters is part of the Gemini/HST Galaxy
Cluster Project, which is described in detail in J\o rgensen et al.\ (2005).
We adopt a $\Lambda$CDM cosmology with $H_0 = 70\,{\rm km\,s^{-1}\,Mpc^{-1}}$, 
$\Omega_M=0.3$, and $\Omega_{\rm \Lambda}=0.7$.

\section{Observational data}

Spectroscopy for RXJ0152.7--1357 and RXJ1226.9+3332 were obtained with
the Gemini Multi-Object Spectrograph (GMOS-N, Hook et al.\ 2004) at Gemini North.
The data for RXJ0152.7--1357 are published in J\o rgensen et al.\ (2005).
The reduction of the RXJ1226.9+3332 spectroscopy was done using similar techniques,
with suitable
changes to take into account the use of the nod-and-shuffle mode of GMOS-N
(J\o rgensen et al.\ in prep.). 
We use Hubble Space Telescope (HST) archive data of the two 
clusters obtained with the Advanced Camera for Surveys (ACS). 
In this paper we use effective radii, $r_{\rm e}$, and mean surface brightnesses, 
$\left < I \right > _{\rm e}$, derived 
from either F775W or F814W observations, calibrated to restframe $B$-band,
see Chiboucas et al.\ (in prep.) for details.
The GALFIT program (Peng et al.\ 2002) was used to determine $r_{\rm e}$ and 
$\left < I \right > _{\rm e}$.
We fit the cluster members with S\'{e}rsic (1968) and $r^{1/4}$ profiles. 
The combination which enters the FP, 
$\log r_{\rm e} + \beta \log \left < I \right > _{\rm e}$ 
($\beta = $0.7-0.8), differs very little for the two choices of profiles.
In the following we use the parameters from $r^{1/4}$-fits for
consistency with our low redshift comparison data. 
None of the main conclusions of this paper would change had
we chosen to use the S\'{e}rsic fits.
Masses of the galaxies are derived as $\rm {\it Mass} = 5 \sigma ^2 r_{\rm e}\,G^{-1}$.

\begin{figure*}
\plotone{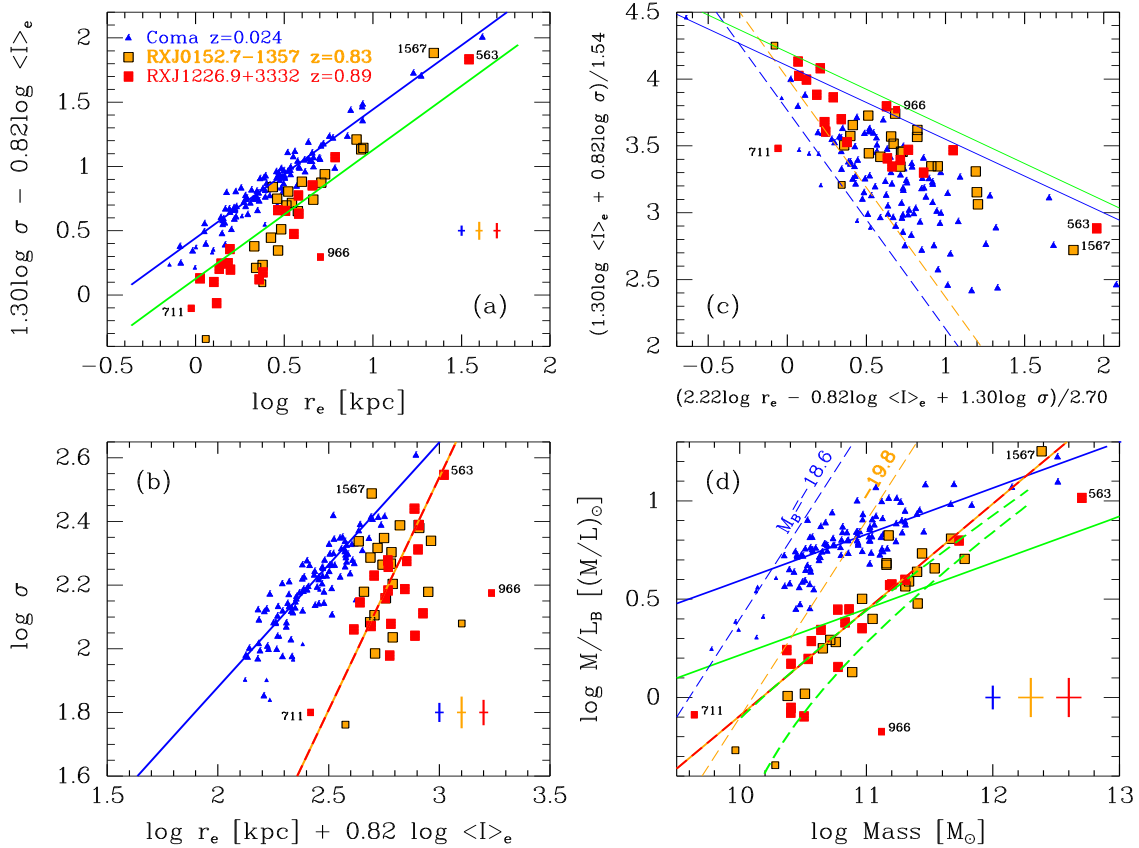}
\caption[]{The FP for RXJ0152.7-0152 (orange), RXJ1226.9+3332 (red),
and Coma (blue). Smaller symbols -- galaxies with ${\it Mass} < 10^{10.3} M_{\sun}$,
excluded from the analysis. 
RXJ1226.9+3332 id=711 and id=966 (with S\'{e}rsic index $n<1.5$) are labeled and excluded 
from the analysis. 
(a) \& (b): Edge-on view of the FP. (c): The FP face-on,
for the Coma cluster coefficients. (d): The FP as Mass {\it vs.} M/L ratio.
Solid blue line on (a), (b) \& (d): Fit to the Coma cluster sample.
Solid green line on (a) \& (d): The Coma cluster fit offset to the median zero
point of the high redshift sample. Orange-red line on (b) \& (d): Fit to
the high redshift sample. The fit shown on (b) is not the optimal FP for the
high redshift sample, since it has the coefficient for $\log \left < I \right >_{\rm e}$ fixed at 0.82.
Dashed lines on (c) and (d): Luminosity
limits for the Coma cluster (blue), and both redshift clusters (orange).
On (c) the solid blue and green lines mark the 
``exclusion zones'' (Bender et al.\ 1992) for the Coma cluster and high redshift
sample, respectively, assuming the slope and zero points as shown on (a).
Dashed green lines on (d): Models from Thomas et al.\ (2005), see text for 
discussion.
Internal uncertainties are shown as representative error bars. On (c) the internal
uncertainties are the size of the points.
\label{fig-FP} }
\end{figure*}

\begin{deluxetable}{lrrrrr}
\tablecaption{Galaxy clusters and samples \label{tab-clusters} }
\tablewidth{0pc}
\tablehead{
\colhead{Cluster} & \colhead{Redshift} & \colhead{$\sigma _{\rm cluster}$\tablenotemark{a}} &
\colhead{N$_{\rm galaxies}\tablenotemark{b}$} & \colhead{N$_{\rm analysis}\tablenotemark{c}$} & \colhead{Ref.\tablenotemark{d}} }
\startdata
Coma=Abell1656  & 0.024 & 1010 $\rm km\,s^{-1}$ & 116 & 105 & (1) \\
RXJ0152.7--1357 & 0.835 & 1110 $\rm km\,s^{-1}$ & 29 & 20 & (2) \\
RXJ1226.7+3332  & 0.892 & 1270 $\rm km\,s^{-1}$ & 25 & 18 & (3) \\
\enddata
\tablenotetext{a}{Cluster velocity dispersion}
\tablenotetext{b}{Number of galaxies observed}
\tablenotetext{c}{Number of galaxies included in the analysis, see text.}
\tablenotetext{d}{(1) J\o rgensen 1999; (2) J\o rgensen et al.\ 2005;
(3) This paper}
\end{deluxetable}

Our Coma cluster sample serves as the low redshift reference sample
(J\o rgensen 1999).
We have obtained new $B$-band photometry of this sample
with the McDonald Observatory 0.8-meter
telescope and the Primary Focus Camera (Claver 1995). The data were reduced in a 
standard fashion and effective parameters were derived as
described in J\o rgensen et al.\ (1995).
Table \ref{tab-clusters} summarizes the sample sizes and some key 
cluster properties.

\section{The Fundamental Plane at z=0.8-0.9}

We first establish the FP for the Coma cluster data. In order to limit
the effect of differences in sample selection for the Coma cluster 
sample and the high redshift sample, we exclude galaxies with 
${\it Mass} < 10^{10.3} M_{\sun}$ as well as emission line galaxies. 
The sum of the absolute residuals perpendicular to the relation was
minimized. We find
\begin{equation}
\log r_{\rm e} = (1.30\pm 0.08) \log \sigma - (0.82\pm 0.03) \log \left < I \right >_{\rm e} - 0.443
\label{eq-Coma}
\end{equation}
where $r_{\rm e}$ is the the effective radius in kpc, $\sigma$ the velocity
dispersion in $\rm km\,s^{-1}$, and $\left < I \right >_{\rm e}$ is the surface brightness
within $r_{\rm e}$ in $L_{\sun}\,{\rm pc^{-2}}$. The uncertainties on the coefficients
are determined using a bootstrap method, see JFK1996 for details.
The rms of the fit is 0.08 in $\log r_{\rm e}$.
The coefficients are in agreement with other determinations available in
the literature (e.g., JFK1996; Colless et al.\ 2001; Blakeslee et al.\ 2002; 
Bernardi et al.\ 2003). 

Figure \ref{fig-FP} shows the Coma cluster FP face-on as well as two edge-on views
of the relation, with the high redshift sample overplotted. 
The FP for the
high redshift sample is not only offset from the Coma cluster FP, but 
appears ``steeper''.
As there is no significant FP zero point difference between the two high redshift 
clusters we treat the high redshift galaxies as one sample.
Deriving the FP for the high redshift sample using the same technique and
sample criteria as for the Coma cluster, we find
\begin{equation}
\log r_{\rm e} = (0.60\pm 0.22) \log \sigma - (0.70\pm 0.06) \log \left < I \right > _{\rm e} +1.13 
\label{eq-highz}
\end{equation}
with an rms of 0.09 in $\log r_{\rm e}$. The difference in the
coefficient for $\log \sigma$ between Eq.\ \ref{eq-Coma} and Eq.\ \ref{eq-highz}
is  $\Delta \alpha = 0.70 \pm 0.23$, a $3\,\sigma$ detection of a difference 
in the FP slope. 
The internal scatter of the two relations is similar.
Figure \ref{fig-FP}d shows the FP as a relation between the galaxy masses
and the M/L ratios.  
The fit to the Coma sample, excluding the low mass galaxies, gives
\begin{equation}
\log {\it M/L}= (0.24\pm 0.03) \log {\it Mass} -1.75
\label{eq-lowzml}
\end{equation}
with an rms of 0.09 in $\log {\it M/L}$.
Fitting the high redshift sample, using the same mass limit, gives
\begin{equation}
\log {\it M/L}= (0.54\pm 0.08) \log {\it Mass} - 5.47
\label{eq-highzml}
\end{equation}
with an rms of 0.14 in $\log {\it M/L}$. The internal scatter in $\log M/L$ 
of the two relations are not significantly different. We find 0.07 and 0.08 
for the Coma sample and the high redshift sample, respectively. 
Even with the same mass limit enforced on both samples one might argue that
the fits are still affected by the difference in the luminosity limit.
Therefore, we also fit a sub-sample of the Coma sample limited at 
$M_B=-19.8\,{\rm mag}$. The coefficient for $\log {\it Mass}$ is in
this case $0.28\pm 0.06$. Thus, the difference between the coefficients
for the high redshift and the low redshift samples is at the $3\,\sigma$ level.

\section{Possible systematic effects}

To test how well we recover input $r_{\rm e}$, $\left < I \right >_{\rm e}$ and 
$\log r_{\rm e} + \beta \log \left < I \right >_{\rm e}$ ($\beta =$0.7-0.8),
we simulate HST/ACS observations of galaxies with
S\'{e}rsic profiles with $n=0.8-4.6$ and effective parameters matching our 
Coma sample.
For $n>2$, the $r^{1/4}$-fits recover $\log r_{\rm e}$ with an rms of 0.15.
However, $\log r_{\rm e} + \beta \log \left < I \right >_{\rm e}$ is recovered with 
an rms scatter of only $\approx 0.02$ for $\beta$ between 0.7 and 0.8.
There are no systematic effects as a function of effective radii or 
luminosities, see Chiboucas et al.\ (in prep.) for details.
Simulations of spectra matching the instrumental resolution, signal-to-noise
ratios and spectral properties of our observational data showed that velocity dispersions
below the instrumental resolution ($\log \sigma = 2.06$) may be subject to 
systematic errors as large as $\pm 0.15$ in $\log \sigma$ (J\o rgensen et al.\ 2005).
Excluding from the 
analysis the four galaxies in the high redshift sample with $\log \sigma < 2.06$, 
we find a slope for the M/L ratio--mass relation of
$0.47\pm 0.06$, while the FP coefficients are not significantly different
from those given in Eq.\ \ref{eq-highz}.

\begin{figure}
\plotone{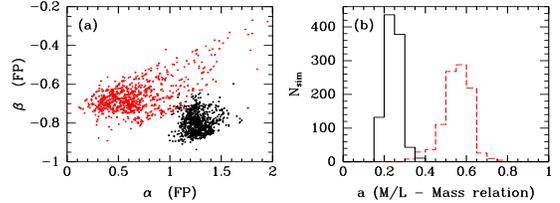}
\caption[]{Distributions of FP coefficients and the slope, $a$, of the M/L ratio--mass relation
for 1000 sub-samples of the Coma cluster sample (black) and for 1000 bootstrap samples
of the high redshift sample (red, dashed). See text for discussion.
\label{fig-sim} }
\end{figure}

Finally, we address whether selection effects can be the cause of the differences
in the relations for the two samples. 
We choose 1000 random sub-samples of 38 galaxies from the Coma sample,
roughly matching the mass distribution of the high redshift sample.
We confirm the match in mass distributions by using a Kolmogorov-Smirnov test. 
The probability that the sub-samples and the real high redshift sample are drawn 
from the same parent distribution is above 90\% for more than 90\% 
of the realizations. For the remainder the probability is above 70\%.
We then compare the fits to these sub-samples to the results from bootstrapping
the high redshift sample. 
For the FP coefficients the sub-sample fits overlap the bootstrap fits in only 1.6\%
of the cases (Fig.\ \ref{fig-sim}a), while for the M/L ratio--mass relation
the slope for the sub-samples overlap with the bootstrap fits in 3.7\%
of the cases (Fig.\ \ref{fig-sim}b).
This shows that the FP and the M/L ratio--mass relation for the high redshift
sample are different from the relations found for the Coma sample
at the 96--98\% confidence level.

Based on the simulations of the data and the selection effects, we conclude 
that the differences in relations we find between the Coma sample 
and the high redshift sample are unlikely to be due to systematic effects
in the data or due to differences in selection effects.

\section{The star formation history of E/S0 cluster galaxies}

The median offset of $\log M/L$ for the high redshift sample relative
to the Coma sample is --0.38.
Using stellar population models from Maraston (2005), which show
that $\Delta \log\,{\it M/L} = 0.935 \Delta \log\,{\it age}$ 
(J\o rgensen et al.\ 2005), this gives an
epoch for the last major star formation episode of $z_{\rm form} \approx 3.2$.
However, the steeper M/L ratio--mass relation found for
high redshift clusters compared to the Coma cluster may be due
to a difference in the epoch of the last major star formation episode
as a function of galaxy mass. The low mass galaxies have experienced
the last major star formation episode much more recently than is
the case for the high mass galaxies.
The difference between the high and low redshift samples is 
$\Delta \log\,{\it M/L} = -0.30 \log\,{\it Mass} + 3.72$,
equivalent to
$\Delta \log\,{\it age} = -0.32 \log\,{\it Mass} + 4.0$.
Thus, for the lowest mass galaxies ($10^{10.3}\,M_{\sun}$) the last epoch 
of star formation may have been as recent as $z_{\rm form} \approx 1.35$.
This is only $\approx$1.5 Gyr prior to when the
light that we now observe was emitted from the galaxies in the high redshift sample. 
There appears to be just enough time for the galaxies to no longer have detectable 
emission lines due to the massive stars formed at that time.
Very shortly after the end of the last major star formation 
episode these galaxies follow a tight FP.
For galaxies with ${\it Mass} \approx 10^{10.8}\,M_{\sun}$ we find
$z_{\rm form} \approx 1.9$, 
while $z_{\rm form} > 4.5$ for galaxies with ${\it Mass} > 10^{11.3}\,M_{\sun}$.

Thomas et al.\ (2005) used absorption line index
data for nearby E/S0 galaxies to establish rough star formation
histories of the galaxies as a function of their masses.
They find that the most massive galaxies form the majority of their
stars at high redshift, while lower mass galaxies continue
forming stars at much later epochs. 
Thomas et al.\ convert velocity dispersions to galaxy masses using
a model dependent relation that is inconsistent with our data. We therefore
correct their masses to consistency with our data by using the 
empirical relation between our mass estimates and the measured 
velocity dispersions.
The lower of the two dashed green lines on Figure \ref{fig-FP}d shows
the result based on the star formation history in high density environments 
as established by Thomas et al.\ and the M/L modeling from Maraston (2005).
Our data show slightly less evolution in the M/L ratios between $z\approx$0.8-0.9
and the present than predicted by Thomas et al. 
However, it is striking that the slope of the predicted relation 
is in agreement with our data.
As an experiment we shifted the predictions from Thomas et al.\ to
the best agreement with our data.
The upper of the two dashed green lines show this for the formation look
back times shifted 2.5 Gyr earlier for all masses such
that the earliest formation look back time is 14 Gyr (roughly the
age of the Universe in this cosmology).
The absolute formation epochs from Thomas et al.\ may not be correct, since their
analysis depends on stellar population models. 
However, their results on the relative timing of the star formation episodes as a function
of galaxy mass closely match our results for this high redshift sample.

Thomas et al.\ predict that star formation is on-going for 
a longer period in low mass galaxies than in high mass galaxies.
Based on this, we estimate that the internal scatter in the M/L--mass 
relation, in $\log M/L$, should be $\approx 0.06$ at $10^{10.3}\,M_{\sun}$ but only
$\approx 0.01$ at $10^{11.3}\,M_{\sun}$. We cannot confirm such a decrease 
of the internal scatter.
However, it would most likely require a larger sample and/or significantly
smaller measurement uncertainties to test this prediction.

Factors other than the mean ages 
of the stellar populations could be affecting the M/L ratios of the 
galaxies. For RXJ0152.7--1357 we found based on absorption line index
data that a large fraction of the galaxies may have $\alpha$-element 
abundance ratios, $\rm [\alpha /Fe]$, about 0.2 dex higher than found in 
nearby clusters (J\o rgensen et al.\ 2005).
This could affect the M/L ratios in a systematic way. Maraston (private 
comm.)\ finds from modeling that stellar populations with
$\rm [\alpha /Fe] = 0.3$, solar metallicities and ages of 2-7 Gyr
may have M/L ratios in the blue that are about 20 per cent higher 
than those with $\rm [\alpha /Fe] = 0.0$. While it is still too early to
use these models for detailed analysis of high redshift data, it 
indicates that for future detailed analysis of the FP we may have
to include information about $\rm [\alpha /Fe]$ of the galaxies.

\section{Conclusions}

We find that the FP for E/S0 galaxies in the clusters 
RXJ0152.7--1357 ($z=0.83$) and RXJ1226.9+3332 ($z=0.89$)
is offset and rotated relative to the
FP of our low redshift comparison sample of Coma cluster galaxies. 
Expressed as a relation between the M/L ratios and the 
masses of the galaxies, the high redshift galaxies follow a significantly 
steeper relation than found for the Coma cluster. We interpret this as
due to a mass dependency of the epoch of the last major star formation
episode. 
The lowest mass galaxies in the sample ($10^{10.3}\,M_{\sun}$) have 
experienced significant star formation as recent as $z_{\rm form} \approx 1.35$, while
high mass galaxies (${\it Mass} > 10^{11.3}\,M_{\sun}$) have $z_{\rm form} > 4.5$.
This is in general agreement with the predictions for the 
star formation histories of E/S0 galaxies from Thomas et al.\ (2005) based 
on their analysis of line index data for nearby galaxies.
The scatter of FP for these two $z=0.8-0.9$ clusters is as low as found
for the Coma cluster, and we find no significant difference in the 
scatter for low and high mass galaxies. This indicates that at a given galaxy mass the 
star formation history for the E/S0 galaxies is quite similar.
In a future paper we will discuss these results in connection with 
our absorption line index data for the galaxies in both high redshift clusters.

\vspace{0.5cm}
Based on observations obtained at the Gemini Observatory (GN-2002B-Q-29, GN-2004A-Q-45), 
which is operated by AURA, Inc., under a cooperative 
agreement with NSF on behalf of the Gemini partnership: NSF
(US), PPARC (UK), NRC (Canada), CONICYT (Chile), 
ARC (Australia), CNPq (Brazil) and CONICET (Argentina).
Based on observations made with the NASA/ESA Hubble Space Telescope. 
IJ, KC, and KF acknowledge support from grant HST-GO-09770.01 from STScI.
STScI is operated by AURA, Inc. under NASA contract NAS 5-26555.


\begin{thebibliography}{}

\bibitem[Bender et al.\(1992)]{bender:1992}
Bender, R., Burstein, D., \& Faber, S. M. 1992, ApJ, 399, 462
 
\bibitem[Bernardi et al.\(2003)]{bernardi:2003}
Bernardi, M., et al. 2003, AJ, 125, 1866

\bibitem[Blakeslee et al.\(2002)]{blakeslee:2002}
Blakeslee, J.\ P., Lucey, J.\ R., Tonry, J.\ L., Hudson, M.\ J.,
Narayanan, V.\ K., \& Barris, B.\ J. 2002, MNRAS, 330, 443

\bibitem[Claver(1995)]{claver:1995}
Claver, C.\ F.\ 1995, Ph.\ D. thesis, Univ. Texas

\bibitem[Colless et al.\(2001)]{colless:2001}
Colless, M., Saglia, R.\ P., Burstein D., Davies, R.\ L., 
McMahan Jr., R.\ K., \& Wegner, G. 2001, MNRAS, 321, 277

\bibitem[de Lucia et al.\(2004)]{lucia:2004}
de Lucia, G., et al. 2004, ApJ, 610, L77

\bibitem[di Serego Alighieri et al.\(2005)]{alighieri:2005}
di Serego Alighieri, S., et al. 2005, A\&A, 442, 125

\bibitem[Djorgovski \& Davis(1987)]{dd:1987}
Djorgovski, S., \& Davis, M. 1987, ApJ, 313, 59

\bibitem[Dressler et al.\(1987)]{dressler:1987}
Dressler, A., Lynden-Bell, D., Burstein, D., Davies, R.\ L., Faber, S.\ M.,
Terlevich, R., Wegner G. 1987, ApJ, 313, 42

\bibitem[Gebhardt et.al.(2003)]{gebhardt:2003}
Gebhardt, K., et al. 2003, ApJ, 597, 239

\bibitem[Holden et al.\(2005)]{holden:2005}
Holden, B.\ P., et al. 2005, ApJ, 620, L83

\bibitem[Hook et al.\(2004)]{hook:2004}
Hook, I.\ M., J{\o}rgensen, I., Allington-Smith, J.\ R., Davies, R.\ L., Metcalfe, N.,
Murowinski, R.\ G., \& Crampton, D. 2004, PASP, 116, 425

\bibitem[J{\o}rgensen(1999)]{IJ:1999}
J{\o}rgensen, I. 1999, MNRAS, 306, 607

\bibitem[J{\o}rgensen et al.(1995)]{IJ:1995} 
J{\o}rgensen I., Franx M., \& Kj{\ae}rgaard, P. 1995, MNRAS, 273, 1097

\bibitem[J{\o}rgensen et al.(1996)]{IJ:1996} 
J{\o}rgensen, I., Franx, M., \& Kj{\ae}rgaard, P. 1996, MNRAS, 280, 167 
(JFK1996)

\bibitem[J{\o}rgensen et al.(1999)]{IJetal:1999}
J{\o}rgensen, I., Franx, M., Hjorth, J., \& van Dokkum, P.\ G. 1999, MNRAS, 308, 833

\bibitem[J{\o}rgensen et al.(2005)]{IJetal:2005}
J{\o}rgensen, I., Bergmann, M., Davies, R., Jordi, B., Takamiya, M., \& Crampton, D. 
2005, AJ, 129, 1249

\bibitem[Kelson et al.(2000)]{kelson:2000}
Kelson, D.\ D., Illingworth, G.\ D., van Dokkum, P.\ G., \& Franx, M. 2000, ApJ, 531, 184

\bibitem[Maraston (2005)]{maraston:2005}
Maraston, C. 2005, MNRAS, 362, 799

\bibitem[Peng et al.(2002)]{peng:2002}
Peng, C.\ Y., Ho, L.\ C., Impey, C.\ D., \& Rix, H.-W. 2002, AJ, 124, 266

\bibitem[S\'{e}rsic(1968)]{sersic:1968}
S\'{e}rsic, J.\ L. 1968, Atlas de Galaxias Australes (C\'{o}rdoba: Obs.\ Astron.\ Univ.\ Nac.\ C\'{o}rdoba)

\bibitem[Thomas et al.\(2005)]{thomas:2005}
Thomas, D., Maraston, C., Bender, R., \& de Oliveira, C.\ M. 2005, ApJ, 621, 673

\bibitem[Toft et al.\(2004)]{toft:2004}
Toft, S., Mainieri, V., Rosati, P., Lidman, C., Demarco, R., Nonino, M., \& Stanford, S.\ A.
2004, A\&A, 422, 29

\bibitem[Treu et al.\(2005)]{treu:2005}
Treu, T., et al. 2005, ApJ, 633, 174 

\bibitem[van der Wel et al.\(2005)]{vanderwel:2005}
van der Wel, A., Franx, M., van Dokkum, P.\ G., Rix, H.-W., Illingworth, G.\ D.,
\& Rosati, P. 2005, ApJ, 631, 145

\bibitem[van de Ven et al.\(2003)]{vandeven:2003}
van de Ven, G., van Dokkum, P.\ G., \& Franx, M. 2003, MNRAS, 344, 924

\bibitem[Wuyts et al.(2004)]{wuyts:2004}
Wuyts, S., van Dokkum, P.\ G., Kelson, D.\ D., Franx, M., \& Illingworth, G.\ D.
2004, ApJ, 605, 677


\bibitem[Ziegler et al.\(2005)]{ziegler:2005}
Ziegler, B.\ L., Thomas, D., B\"{o}hm, A., Bender, B., Fritz, A., \& Maraston, C.
2005, A\&A, 433, 519

\end{thebibliography}
\end{document}